\documentclass{aa}
\usepackage{graphicx}
\usepackage{txfonts}
\usepackage{natbib}
\bibpunct{(}{)}{;}{a}{}{,}

\begin{document}

\title{Spectral Energy Distributions of 6.7 GHz methanol masers}

\author{J. D. Pandian
  \inst{1,2}\thanks{Current affiliation at IfA, U. Hawaii}
  \and
  E. Momjian
  \inst{3}
  \and
  Y. Xu
  \inst{4}
  \and
  K. M. Menten
  \inst{1}
  \and
  P. F. Goldsmith
  \inst{5}
  }
\institute{
  Max-Planck-Institut f\"{u}r Radioastronomie, Auf dem H\"{u}gel 69, 53121 Bonn, Germany\\
  \email{[jpandian,kmenten]@mpifr-bonn.mpg.de}
  \and
  Institute for Astronomy (IfA), University of Hawaii, 2680 Woodlawn Drive, Honolulu, HI 96814, USA\\
  \email{jpandian@ifa.hawaii.edu}
  \and
  National Radio Astronomy Observatory, P.O. Box O, Socorro, NM 87801, USA\\
  \email{emomjian@aoc.nrao.edu}
  \and
  Purple Mountain Observatory, Chinese Academy of Sciences, Nanjing 210008, China\\
  \email{xuye@pmo.ac.cn}
  \and
  Jet Propulsion Laboratory, California Institute of Technology, Pasadena, CA 91109, USA\\
  \email{Paul.F.Goldsmith@jpl.nasa.gov}
  }

\abstract
{The 6.7 GHz maser transition of methanol has been found exclusively towards massive star forming regions. A majority of the masers have been found to lack the presence of any associated radio continuum. This could be due to the maser emission originating prior to the formation of an \ion{H}{ii} region around the central star, or from the central object being too cool to produce a \ion{H}{ii} region.}
{One way to distinguish between the two scenarios is to determine and model the spectral energy distributions (SEDs) of the masers.}
{We observed a sample of 20 6.7 GHz methanol masers selected from the blind Arecibo survey, from centimeter to submillimeter wavelengths. We combined our observations with existing data from various Galactic plane surveys to determine SEDs from centimeter to near-infrared wavelengths.}
{We find that 70\% of the masers do not have any associated radio continuum, with the rest of the sources being associated with hypercompact and ultracompact \ion{H}{ii} regions. Modeling the SEDs shows them to be consistent with rapidly accreting massive stars, with accretion rates well above $10^{-3}$ $M_\odot$ yr$^{-1}$. The upper limits on the radio continuum are also consistent with any ionized region being confined close to the stellar surface.}
{This confirms the paradigm of 6.7 GHz methanol masers being signposts of early phases of massive star formation, mostly prior to the formation of a hypercompact \ion{H}{ii} region.}


\maketitle

\section{Introduction}
Class II methanol masers, of which the 6.7 GHz maser is the strongest, are commonly found towards massive star forming regions. A number of early surveys for 6.7 GHz methanol masers were focused towards IRAS sources with colors suggestive of ultracompact \ion{H}{ii} (UC \ion{H}{ii}) regions (e.g. \citealt{szym00a} and references therein). While methanol masers appeared to be associated with \ion{H}{ii} regions at low angular resolution, subsequent high-resolution observations demonstrated that the masers are generally not coincident with strong radio sources (e.g. \citealt{wals98,phil98}). Further, statistics from blind surveys showed that most methanol masers did not have associated radio continuum, at least not at the few mJy level at a wavelength of 2 cm  -- only 10\% of the sources detected in the Arecibo Methanol Maser Galactic Plane Survey \citep[AMGPS;][]{pand07a} had a counterpart at 21 cm within 23$''$ of the maser position, while two-thirds of the methanol masers in the Toru\'{n} blind survey \citep{szym02} did not have a 6 cm counterpart within 1$'$.

The lack of radio continuum towards methanol masers has been interpreted in two ways -- (i) The masers are associated with massive stars prior to the origin of an ultracompact region \citep{wals98}. Rapid accretion onto a massive star can quench the formation of a \ion{H}{ii} region \citep{walm95}. (ii) Alternately, the exciting star has a mass lower than that of a prototypical high-mass star (spectral type B3 or earlier), and is thus too cool to produce a detectable \ion{H}{ii} region \citep{phil98}. \citet{phil98} concluded the latter based on the expected number of UC \ion{H}{ii} regions in the Galaxy assuming that methanol masers overlapped with UC \ion{H}{ii} regions for only 20\% of their lifetimes. However, subsequent work has favored the first hypothesis. For example, \citet{van03} argue that it is unlikely for the masers to be associated with stars of spectral type B5 based on the expected lifetime of maser emission and the number of methanol masers in the Galaxy. They further argue that the efficiency of exciting a maser would decrease dramatically with decreasing stellar mass.

Surveys for 6.7 GHz methanol masers towards sites of low-mass star formation have not yielded any detections \citep{mini03,xu08}. Based on their survey, \citet{mini03} give a lower limit to the stellar mass associated with bright (brightness temperature $> 3 \times 10^6$ K) 6.7 GHz methanol masers as 3 $M_\odot$. The bolometric luminosities of the sources associated with 6.7 GHz maser emission, estimated using mid-infrared and IRAS data, were suggestive of stars with strong ionizing radiation \citep{wals99}. \citet{wals01} modeled the spectral energy distributions of the masers using IRAS and near-infrared data and found that sources for which spectral types of the central stars could be unambiguously identified were consistent with massive, ionizing stars. \citet{mini05} carried out a multi-wavelength study of five radio-quiet methanol masers, and found the environment around the masers to be characteristic of massive star forming clumps in earlier evolutionary phases than \ion{H}{ii} regions. The observation of warm dust around methanol masers with bolometric luminosities greater than 10$^3$ $L_\odot$ \citep{wals03} also suggested that the masers are associated with massive star forming regions.

One of the limitations of these studies has been that the methanol maser sample is not complete. They are typically selected from a targeted survey, and are hence potentially biased. The work of \citet{elli06}, which investigated the mid-infrared colors of the masers using the Galactic Legacy Infrared Mid-Plane Survey Extraordinaire survey (GLIMPSE; \citealt{benj03}), did use a complete sample, but no modeling of the spectral energy distributions (SEDs) was carried out on account of polycyclic aromatic hydrocarbon (PAH) emission and silicate absorption lines falling in the Spitzer IRAC bands. 

Recently, a grid of 200,000 young stellar object (YSO) models was developed by \citet{robi06} spanning a wide range of evolutionary stages for different stellar masses to model the SED from optical to millimeter wavelengths. In addition, 24 $\mu$m data is now available for the Galactic plane from the Spitzer legacy project, MIPSGAL \citep{care09}. The 6$''$ beam of Spitzer at this wavelength and its much better sensitivity is a significant improvement to that of the Midcourse Source Experiment satellite (MSX; \citealt{pric99}). In this paper, we present multi-wavelength observations from centimeter to submillimeter wavelengths towards a sample of 20 6.7 GHz methanol masers selected from a blind survey. We combine our observations with existing data to determine SEDs from centimeter to near-infrared wavelengths. The combination of the radio continuum data and the results from SED modeling give very strong evidence that the methanol masers are associated with rapidly accreting massive stars.

\section{Sample, observations and data reduction}
\subsection{The sample}\label{sample}
To determine the spectral energy distributions of 6.7 GHz methanol masers, we observed a sample of 20 masers with the Very Large Array\footnote{The National Radio Astronomy Observatory is a facility of the National Science Foundation operated under cooperative agreement by Associated Universities, Inc.} (VLA) at 3.6 cm, 1.3 cm and 6.9 mm, the IRAM 30-meter telescope at 1.2 mm and the Atacama Path Finder Experiment telescope, APEX\footnote{This publication is based on data acquired with the Atacama Pathfinder Experiment (APEX). APEX is a collaboration between the Max-Planck-Institut f\"{u}r Radioastronomie, the European Southern Observatory, and the Onsala Space Observatory.}\citep{gues06}, at 870 $\mu$m. The source sample was selected from the AMGPS and included all sources detected between Galactic longitudes of $38.6^\circ$ and $43.1^\circ$, and Galactic latitudes $|b| \leq 0.42^\circ$, the latitude constraint being imposed for the sake of completeness. The peak flux density of the masers range from 0.2 Jy to 26 Jy, and their luminosities \citep[taken from][]{pand09} range from $9.9 \times 10^{-8}$ to $2.6 \times 10^{-5}~L_\odot$.

14 out of 20 sources have accurate positions (astrometry better than 0.1$''$) determined from phase-referenced observations with the MERLIN interferometer (these observations were part of a larger project to determine accurate positions to the AMGPS sources and will be described in detail in Pandian et al. 2010, in preparation). The remaining sources have positions determined from 24 $\mu$m point sources in MIPSGAL. It is generally found that 6.7 GHz methanol masers coincide with 24 $\mu$m point sources with good accuracy (\citealt{xu09}, Pandian et al., in preparation). For instance, the mean deviation between accurate maser positions and their corresponding 24 $\mu$m MIPSGAL point sources for the 14 sources above is 1.1$''$. Consequently, for sources without good interferometer positions, we can use MIPSGAL point sources to determine positions accurate to $\sim 1''$.

As indicated in \citet{pand07b}, the AMGPS positions have a root mean square positional accuracy to $7''$, which translates to an uncertainty of 18$''$ at the 95\% confidence level. Hence, we looked for MIPSGAL point sources within 18$''$ of the 6 methanol masers without good interferometer positions. In most cases, there was a single point source within the search radius that was defined as the counterpart for the maser. In cases where there were more than one 24 $\mu$m source, we looked at mid-infrared colors using the GLIMPSE point source catalog, and selected counterparts that had colors consistent with those of young stellar objects. 

The positions, peak flux densities, distances and luminosities of the masers in our sample are shown in Table 1.

\begin{table*}
\caption{6.7 GHz methanol masers observed at multiple wavelengths to determine SEDs. The columns show the source name, equatorial J2000 coordinates, peak flux density, distance and luminosity. Positions that are accurate to 0.1$''$ are derived from MERLIN observations, while positions accurate to 1$''$ are derived from MIPSGAL 24 $\mu$m point sources.}
\label{table1}
\centering
\begin{tabular}{lccccc}
\hline \hline
Source & $\alpha$ & $\delta$ & $S_p$ & $d$ & $L_m$ \\
       & (J2000)  & (J2000)  & (Jy)  & (kpc) & ($L_\odot$) \\
\hline
38.66+0.08 & 19 01 35.24 & 05 07 47.4 & 0.71 & $16.3 \pm 0.9$ & $(1.3 \pm 0.2) \times 10^{-6}$ \\
38.92-0.36 & 19 03 38.66 & 05 09 42.5 &  1.29 & $10.5 \pm 0.4$ & $(9.9 \pm 0.8) \times 10^{-7}$ \\
39.39-0.14 & 19 03 45.32 & 05 40 42.6 &  0.77 & $4.3 \pm 0.5$  & $(9.9 \pm 2.3) \times 10^{-8}$ \\
39.54-0.38 & 19 04 52.6  & 05 42 08   &  0.24 & $9.0 \pm 0.5$  & $(1.3 \pm 0.2) \times 10^{-7}$ \\
40.28-0.22 & 19 05 41.22 & 06 26 12.7 & 68.03 & $4.9^{+0.9}_{-0.6}$ & $1.1^{+0.5}_{-0.2} \times 10^{-5}$ \\
40.62-0.14 & 19 06 01.59 & 06 46 36.1 &  6.97 & $10.5 \pm 0.4$ & $(5.3 \pm 0.4) \times 10^{-6}$ \\
40.94-0.04 & 19 06 15.38 & 07 05 54.5 &  1.50 & $10.0 \pm 0.5$ & $(1.0 \pm 0.1) \times 10^{-6}$ \\
41.08-0.13\tablefootmark{a} & 19 06 48.5  & 07 11 00   &  0.33 & $8.4 \pm 0.6$  & $(1.6 \pm 0.3) \times 10^{-7}$ \\
41.12-0.11 & 19 06 50.2  & 07 14 02   &  0.63 & $10.0 \pm 0.5$ & $(4.4 \pm 0.5) \times 10^{-7}$ \\
41.12-0.22 & 19 07 14.86 & 07 11 00.7 &  1.23 & $8.7 \pm 0.6$  & $(6.5 \pm 0.9) \times 10^{-7}$ \\
41.16-0.20 & 19 07 14.3  & 07 13 19   &  0.20 & $8.7 \pm 0.6$  & $(1.0 \pm 0.2) \times 10^{-7}$ \\
41.23-0.20 & 19 07 21.38 & 07 17 08.2 &  7.55 & $8.7 \pm 0.5$  & $(4.0 \pm 0.5) \times 10^{-6}$ \\
41.27+0.37 & 19 05 23.6  & 07 35 06   &  0.16 & $11.5 \pm 0.5$ & $(1.5 \pm 0.2) \times 10^{-7}$ \\
41.34-0.14 & 19 07 21.84 & 07 25 17.3 & 25.40 & $11.6 \pm 0.5$ & $(2.4 \pm 0.2) \times 10^{-5}$ \\
41.58+0.04 & 19 07 09.1  & 07 42 25   &  0.22 & $11.5 \pm 0.5$ & $(2.0 \pm 0.2) \times 10^{-7}$ \\
42.03+0.19 & 19 07 28.18 & 08 10 53.5 & 30.10 & $11.1 \pm 0.5$ & $(2.6 \pm 0.3) \times 10^{-5}$ \\
42.30-0.30 & 19 09 43.59 & 08 11 41.4 &  5.07 & $10.5 \pm 0.5$ & $(3.9 \pm 0.4) \times 10^{-6}$ \\
42.43-0.26 & 19 09 49.86 & 08 19 45.4 &  1.61 & $7.9 \pm 0.8$  & $(7.0 \pm 1.4) \times 10^{-7}$ \\
42.70-0.15 & 19 09 55.07 & 08 36 53.5 &  4.03 & $15.9 \pm 0.8$ & $(7.1 \pm 0.7) \times 10^{-6}$ \\
43.08-0.08 & 19 10 22.05 & 08 58 51.5 &  4.11 & $11.2 \pm 0.5$ & $(3.6 \pm 0.3) \times 10^{-6}$ \\
\hline
\end{tabular}
\tablefoot{
\tablefoottext{a}{An alternative position for this source based on MIPSGAL is (19$^h$ 06$^m$ 49$^s$.0, 7$^\circ$ 11$'$ 07$''$).}
}
\end{table*}

\subsection{Observations and data reduction}
The 3.6 cm observations were carried out on December 8, 2007 with the VLA in the B-configuration. The correlator was set up in the standard continuum mode, and provided a total bandwidth of 100 MHz per polarization. The integration time was 2.5 minutes per source. The source 3C286 was used as a primary and flux calibrator while J1856+061 served as the phase calibrator. The data were reduced using standard procedures in the Astronomical Image Processing System (AIPS) of NRAO. The full width at half-maximum (FWHM) of the synthesized beam was $0.9'' \times 0.8''$, and the $1\sigma$ noise in the images range from 0.13 to 0.24 mJy/beam.

The 1.3 cm observations were carried out with the VLA on March 22, 2008 in the C-configuration. The source 3C286 was used as a primary flux calibrator while J1851+005 served as the phase calibrator. The observations were carried out using fast switching for good phase stability. Each fast switching cycle comprised of 140 s on the target followed by 40 s on the phase calibrator. The total on-source integration time was 6 to 13 minutes per source. The data were reduced using AIPS. The $1\sigma$ noise in the maps was around 0.12 mJy/beam, the synthesized beam being $1.0'' \times 0.9''$.

A subset of the sources in the sample (see Table \ref{table2}) were observed with the VLA at 6.9 mm in the D-configuration on September 2, 2008. As with the 1.3 cm observations, 3C286 was used as a primary flux calibrator while J1851+005 served as the phase calibrator. Fast switching was used, each cycle comprising of 240 s on the target followed by 60 s on the phase calibrator, with the total on-source integration times ranging from 4 to 31 minutes depending on the strength of the source. The data were reduced using AIPS. The root mean square noise level in the maps ranged from 0.4 to 1.7 mJy/beam and the FWHM synthesized beam sizes ranged from $2.7'' \times 2.2''$ to $3.8'' \times 2.3''$.

The 1.2 mm observations were carried out between January and March 2008 as part of ``pooled observations'' using the 117 element MAMBO-II bolometer camera \citep{krey98} at the IRAM 30-meter telescope. Each source was mapped using the standard on-the-fly technique such that a $\sim 2'$ region was uniformly sampled (although the actual extents of the maps are considerably larger). The time spent per map was around 12 minutes with pointing, focus and calibration scans being done at regular intervals. The data were reduced using the ``mapCSF'' procedure in the MOPSIC, which is part of the GILDAS software package (http://www.iram.fr/IRAMFR/GILDAS). The root mean square level in the maps was typically 8 to 10 mJy/beam with the FWHM beam size being $\sim 10.5''$. The maps were loaded into AIPS, and flux densities of point sources were extracted using the task ``JMFIT''.

The 870 $\mu$m observations were carried out in December 2007 using the Large APEX Bolometer Camera (LABOCA; \citealt{siri09}) at the APEX telescope. The field of view of the camera ($\sim 11'$) was mapped around each source (except for sources 41.08--0.13 and 41.12--0.11, and 41.12--0.22, 41.16--0.20 and 41.23--0.20, which are close enough to be covered in two separate fields of view) using raster scans. The time spent per map ranged from 5 to 22 minutes depending on the strength of the source and the weather conditions. The data were reduced using the Bolometer array data Analysis (BoA; Schuller et al., in prep.) software. The data reduction strategy was similar to that described by \citet{schul09} and involved flagging of bad pixels, removal of correlated noise, despiking, low-frequency filtering and first-order baselining of the timestream followed by building a map using appropriate weighting. These steps were carried out in several iterations, with a source model from the previous iteration being subtracted from the data, enabling more aggressive application of procedures such as despiking. The final $1\sigma$ noise level in the maps ranged from 30 to 75 mJy/beam, the FWHM beam size being $\sim 18.2''$. As with the MAMBO data, flux densities of point sources were extracted using the AIPS task ``JMFIT''.

In addition to our observations above, we also used existing data at mid-infrared wavelengths to determine the SEDs. While the point source catalogs of the GLIMPSE and 2MASS surveys provided data from 8.0 to 1.24 $\mu$m, the point source catalog of 24 $\mu$m sources from the MIPSGAL survey has not been released. Hence, we determined 24 $\mu$m flux densities directly from the mosaics released by the MIPSGAL team. To measure the flux densities, we followed the procedure outlined in the cookbook in the Spitzer Science Center website (http://ssc.spitzer.caltech.edu/dataanalysistools/cookbook/25/). For sources that were saturated in the MIPS images, we used flux densities from the Galactic Plane survey carried out with the MSX satellite \citep{egan03}.

\section{Results}

No continuum emission was seen in 14 out of 20 sources at both 3.6 cm and 1.3 cm. At the native resolution, six sources were detected at 1.3 cm, of which five were detected at 6.9 mm and two were detected at 3.6 cm and at 6 cm in the CORNISH survey \citep{purc08}. The latter two sources have spectral indices consistent with optically thin free-free emission, while the three sources detected at 6.9 mm have spectral indices greater than 0.3, which is consistent with optically thick free-free emission. The nature of the source 42.70--0.15, that was detected only at 1.3 cm is not clear. However, its $3\sigma$ flux density limit at 3.6 cm would be consistent with optically thick emission between these wavelengths.

One of the concerns in the above analysis is the presence of extended emission that is resolved in our observations. Several UC \ion{H}{ii} regions have been observed to be embedded in extended radio continuum emission \citep{kurt99}. The recent work of \citet{long09} also found a significant fraction of 6.7 GHz methanol maser sources that had no 3.6 cm radio continuum as measured by \citet{wals98}, to have extended emission with the total flux at times being several orders of magnitude discrepant from previous limits. Missing an extended radio continuum component due to spatial filtering of an interferometer can lead to errors in the classification and inferred age of the source. 

We hence re-imaged all sources at 1.3 cm by tapering the uv-coverage of the data. This decreases the angular resolution of the data (at the expense of slightly increased noise in the images) enabling identification of extended emission. The 1.3 cm observations are sensitive to a largest angular scale of $\sim 30''$\footnote{http://www.vla.nrao.edu/astro/guides/vlas/current/node10.html}. With the exception of 39.39--0.14, no extended emission was found in the sources. Although this does not rule out the presence of extended emission on scales larger than $30''$, it suggests a lack of continuum sources similar to those discovered by \citet{long09}. In the case of 39.39--0.14, the extended emission led to the discovery of a counterpart at 3.6 cm leading to re-classification of the nature of radio continuum emission from optically thick to optically thin emission between 3.6 cm and 1.3 cm. A more detailed discussion of this source can be found in \S\ref{indi}.

\begin{table*}
\caption{Centimeter to submillimeter wave flux densities measured towards the 20 6.7 GHz methanol masers. The indicated upper limits are $3\sigma$ and all flux densities are in mJy.}
\label{table2}
\centering
\begin{tabular}{lccccc}
\hline \hline
Source & $S$(3.6 cm) & $S$(1.3 cm) & $S$(6.9 mm) & $S$(1.2 mm) & $S$(870 $\mu$m) \\
\hline
38.66+0.08 & $14.6 \pm 0.3$ & $11.7 \pm 0.2$ & $11.9 \pm 3.6$ & \tablefootmark{a} & $1390 \pm 320$ \\
38.92-0.36 & $< 0.6$ & $< 0.3$ & \ldots & $1210 \pm 130$ & \tablefootmark{b} \\
39.39-0.14 & $3.5 \pm 1.4$ & $3.0 \pm 0.5$ & $5.0 \pm 1.3$ & $1514 \pm 48$ & $4550 \pm 170$ \\
39.54-0.38 & $< 0.7$ & $< 0.3$ & $< 1.8$ &  $154 \pm 62$ & $480 \pm 120$ \\
40.28-0.22 & $< 0.8$ & $0.4 \pm 0.2$ & $6.0 \pm 1.2$ &  $3156 \pm 35$ & $12520 \pm 160$ \\
40.62-0.14 & $< 0.8$ & $0.9 \pm 0.3$ & $3.9 \pm 1.7$ &  $1657 \pm 56$ & $6280 \pm 280$ \\
40.94-0.04 & $< 0.7$ & $< 0.3$ & \ldots &  $53 \pm 26$ & $170 \pm 110$ \\
41.08-0.13 & $< 0.6$ & $< 0.4$ & \ldots &  $680 \pm 120$ & $2670 \pm 300$ \\
41.12-0.11 & $< 0.5$ & $< 0.4$ & \ldots &  $390 \pm 52$ & $1360 \pm 230$ \\
41.12-0.22 & $< 0.6$ & $< 0.4$ & \ldots &  $367 \pm 55$ & $3130 \pm 310$ \\
41.16-0.20 & $< 0.5$ & $< 0.4$ & \ldots & \tablefootmark{c} & \tablefootmark{c} \\
41.23-0.20 & $< 0.5$ & $< 0.4$ & \ldots &  $264 \pm 35$ & $1590 \pm 250$ \\
41.27+0.37 & $< 0.5$ & $< 0.4$ & \ldots &  $309 \pm 60$ & $860 \pm 100$ \\
41.34-0.14 & $< 0.5$ & $< 0.4$ & \ldots &  $327 \pm 39$ & $1140 \pm 230$ \\
41.58+0.04 & $< 0.4$ & $< 0.4$ & $< 1.3$ & $< 28$ & $< 110$ \\
42.03+0.19 & $< 0.4$ & $< 0.4$ & \ldots &  $572 \pm 81$ & $1590 \pm 260$ \\
42.30-0.30 & $< 0.4$ & $< 0.4$ & $< 1.5$ &  $538 \pm 57$ & $2610 \pm 200$ \\
42.43-0.26 & $59.5 \pm 1.2$ & $49.4 \pm 0.8$ & $45.9 \pm 5.5$ &  $670 \pm 36$ & $2230 \pm 160$ \\
42.70-0.15 & $< 0.4$ & $0.8 \pm 0.2$ & $< 4.5$ &  $177 \pm 44$ & $500 \pm 130$ \\
43.08-0.08 & $< 1.5$ & $< 0.5$ & \ldots &  $595 \pm 67$ & $2220 \pm 49$ \\
\hline
\end{tabular}
\tablefoot{
\tablefoottext{a}{1.2 mm flux density uncertain due to presence of extended emission that is not fully recovered.} \\
\tablefoottext{b}{Source not detected at 870$\mu$m due to poorer resolution of LABOCA.} \\
\tablefoottext{c}{No point source seen due to source being in the middle of a filament.}
}
\end{table*}

\begin{table*}
\caption{Infrared flux densities of the 20 6.7 GHz methanol masers.}
\label{table3}
\centering
\begin{tabular}{lcccccccccccc}
\hline \hline
Source & MIPS & \multicolumn{4}{c}{MSX} & \multicolumn{4}{c}{IRAC} & \multicolumn{3}{c}{2MASS} \\
 & 24 $\mu$m & 21.3 $\mu$m & 14.7 $\mu$m & 12.1 $\mu$m & 8.3 $\mu$m & 8 $\mu$m & 5.8 $\mu$m & 4.5 $\mu$m & 3.6 $\mu$m & $K$ & $H$ & $J$ \\
 & (Jy) & (Jy) & (Jy) & (Jy) & (Jy) & (mJy) & (mJy) & (mJy) & (mJy) & (mJy) & (mJy) & (mJy) \\
\hline
38.66+0.08 & 2.37 & 2.37 & \ldots & \ldots & 0.14 & \ldots & 19.5 & 7.70 & 0.80 & \ldots & \ldots & \ldots \\
38.92-0.36 & 0.051 & \ldots & \ldots & \ldots & \ldots & \ldots & \ldots & 3.23 & 0.40 & \ldots & \ldots & \ldots \\
39.39-0.14 & 3.88 & \ldots & \ldots & \ldots & \ldots & 113 & 144 & 72.1 & \ldots & \ldots & \ldots & \ldots \\
39.54-0.38 & 0.13 & \ldots & \ldots & \ldots & \ldots & 4.74 & 4.89 & 4.35 & 1.44 & \ldots & \ldots & \ldots \\
40.28-0.22 & $> 5.9$\tablefootmark{a} & 7.15 & 1.03 & 0.064 & \ldots & 289 & 225 & 252 & 129 & 19.2 & 3.70 & \ldots \\
40.62-0.14 & \tablefootmark{a} & 29.1 & 4.94 & 2.29 & 0.98 & 1522 & 962 & 355 & \ldots & 3.03 & \ldots & \ldots \\
40.94-0.04 & 0.085 & \ldots & \ldots & \ldots & \ldots & \ldots & \ldots & 0.87 & 0.42 & \ldots & \ldots & \ldots \\
41.08-0.13A & 0.055 & \ldots & \ldots & \ldots & \ldots & 6.12 & 8.11 & 7.65 & 2.51 & \ldots & \ldots & \ldots \\
41.08-0.13B & 0.081 & \ldots & \ldots & \ldots & \ldots & \ldots & \ldots & 4.22 & 2.38 & 2.52 & 1.56 & \ldots \\
41.12-0.11 & 4.03 & 2.73 & 2.73 & 1.83 & 1.27 & 1482 & 1526 & \ldots & 317 & 6.02 & \ldots & \ldots \\
41.12-0.22A & 0.15 & \ldots & \ldots & \ldots & \ldots & \ldots & \ldots & 3.87 & 2.02 & \ldots & \ldots & \ldots \\
41.12-0.22B & 0.15 & \ldots & \ldots & \ldots & \ldots & \ldots & 1.23 & 1.34 & 0.32 & \ldots & \ldots & \ldots \\
41.16-0.20 & 0.17 & \ldots & \ldots & \ldots & \ldots & \ldots & 1.23 & 1.34 & 0.32 & \ldots & \ldots & \ldots \\
41.23-0.20 & 0.26 & 1.13\tablefootmark{b} & 0.97\tablefootmark{b} & 1.06\tablefootmark{b} & 0.69\tablefootmark{b} & \ldots & \ldots & 8.07 & 4.37 & \ldots & \ldots & \ldots \\
41.27+0.37 & 2.33 & 1.51 & 0.81 & 0.95 & 0.37 & 343 & 183 & 106 & 54.7 & 7.88 & \ldots & \ldots \\
41.34-0.14 & 0.91 & \ldots & \ldots & \ldots & \ldots & 20.5 & 11.9 & 5.59 & \ldots & \ldots & \ldots & \ldots \\
41.58+0.04 & 0.019 & \ldots & \ldots & \ldots & \ldots & \ldots & \ldots & 0.55\tablefootmark{c} & 0.73\tablefootmark{c} & \ldots & \ldots & \ldots \\
42.03+0.19 & $> 5.3$\tablefootmark{a} & 15.96 & 9.40 & 5.97 & 2.42 & 1796 & 2362 & 681 & 463 & 38.8 & 2.84 & \ldots  \\
42.30-0.30 & 2.49 & \ldots & \ldots & \ldots & \ldots & 89.6 & 47.9 & 27.2 & 6.58 & \ldots & \ldots & \ldots \\
42.43-0.26 & \tablefootmark{a} & 67.58 & 18.84 & 12.65 & 4.47 & \ldots & 1118 & 991 & 565 & 210 & 60.0 & 8.01 \\
42.70-0.15 & 2.88 & \ldots & \ldots & \ldots & \ldots & 36.7 & 11.9 & 3.38 & 1.27 & \ldots & \ldots & \ldots \\
43.08-0.08 & 0.83 & \ldots & \ldots & \ldots & \ldots & \ldots & 3.24 & 5.02 & 1.06 & \ldots & \ldots & \ldots \\
\hline
\end{tabular}
\tablefoot{
\tablefoottext{a}{Saturated in MIPS data.}\\
\tablefoottext{b}{Significant uncertainty in the MSX flux densities on account of the extended structure seen in the higher resolution MIPS data.}\\
\tablefoottext{c}{An offset of 2.8$''$ between the GLIMPSE and MIPSGAL point sources casts some doubt about the association of the two sources.}
}
\end{table*}

At 1.2 mm wavelength, 18 out of 20 sources were detected as point sources, of which 17 were also detected at 870 $\mu$m. The 1.2 mm maps of the sources are shown online (Fig. \ref{mambomaps}). One source, 41.16--0.20, was found to be in the middle of a filament at both 1.2 mm and 870 $\mu$m, without any point source to extract flux densities from. All sources had counterparts at 24 $\mu$m (see \S\ref{sample}), and in the GLIMPSE point source catalog with flux densities in at least two IRAC bands. In addition, 7 sources also had counterparts in the 2MASS point source catalog. The flux densities of the sources from centimeter to submillimeter wavelengths are shown in Table \ref{table2}, while the infrared flux densities are shown in Table \ref{table3}.

\subsection{Notes on selected sources}\label{indi}
{\it 38.66+0.08} -- The 3.6 and 1.3 cm images show a prominent point source along with some hint of extended emission. The point source flux density measured with ``JMFIT'' is 14.6, 11.7 and 11.9 mJy respectively at 3.6, 1.3 and 0.69 cm suggestive of a flat spectral index. However, the 6 cm flux density measured by CORNISH is only 6.97 mJy, and the resulting spectral index (defined as $S_\nu \propto \nu^\alpha$ where $S_\nu$ is the flux density at frequency $\nu$ and $\alpha$ is the spectral index) between 6 cm and 3.6 cm is +1.3 which is consistent with optically thick free-free emission. Using the Altenhoff approximation, the free-free optical depth is given by \citep{alte60}
\begin{equation}
\tau_\nu = \frac{0.082}{\nu^{2.1}{T_e^{1.35}}} \int n_e^2~dl
\end{equation}
where $\nu$ is in GHz and $\int n_e^2~dl$, the emission measure, is in pc cm$^{-6}$. Since the free-free emission in this source is optically thick at 6 cm ($\nu$ = 4.86 GHz), but optically thin at 3.6 cm ($\nu$ = 8.46 GHz), the emission measure is between $8.5 \times 10^7$ and $2.7 \times 10^8$ pc cm$^{-6}$, and can be classified as a UC \ion{H}{ii} region. 

It is to be noted that the 1.4 GHz NRAO VLA Sky Survey (NVSS; \citealt{cond98}) reveals a 165 mJy source about 8.5$''$ away from the maser position, with a spatial extent of $63'' \times 58''$. Although this is well above the maximum angular scale that can be recovered by our observations, we do not recover flux above the point source component (which is coincident with the maser position) indicated in Table \ref{table2} when imaging the source with a tapered uv-coverage. Considering that this source is at a distance of 16.3 kpc from the Sun, the NVSS results are indicative of a star forming complex with a classical \ion{H}{ii} region, embedded within which is the UC \ion{H}{ii} region detected in our observations.

The MAMBO observations show a complex structure with significant extended emission (which is consistent with the extent of the NVSS emission) that cannot be fully recovered from the small size of the map, and consequently no 1.2 mm flux density is reported in Table \ref{table2}. However, the LABOCA map does not show any extended emission within the map sensitivity.

\begin{figure}
\centering
\includegraphics[width=0.45\textwidth]{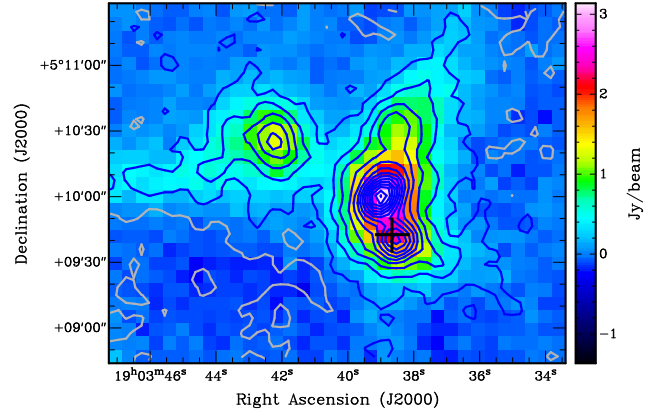}
\caption{The 870 $\mu$m LABOCA image of source 38.92--0.36 is shown in color, overlaid on which is the 1.2 mm MAMBO image in green contours. The contours start at 10\% of the peak emission (550 mJy/beam) in steps of 5\% thereafter. The black cross shows the location of the methanol maser.}
\label{g38.92l}
\end{figure}

{\it 38.92--0.36} -- The LABOCA data shows a strong point source at J2000 coordinates (19$^h$ 38$^m$ 38$^s$.9, 5$^\circ$ 10$'$ 0$''$) with a bow-shock like morphology (Fig. \ref{g38.92l}). No point source is seen at the location of the maser, and consequently, no 870 $\mu$m flux density is reported in Table \ref{table2}. The higher resolution MAMBO data resolves the single LABOCA source into multiple sources including a point source at the maser site.

{\it 39.39--0.14} -- At native resolution, this source appears to be a hypercompact \ion{H}{ii} (HC \ion{H}{ii}) region since there is no counterpart detected at 3.6 cm, while the spectral index between 1.3 cm and 6.9 mm suggests optically thick free-free emission. However, imaging the source with a tapered uv distribution shows the presence of extended emission with a total 1.3 cm flux of 3.0 mJy, the extended component accounting for 50\% of the flux. A 3.6 cm image made with a similar tapered uv-coverage shows a 3.5$\sigma$ point source with a total flux of 3.5 mJy, indicating that the emission is optically thin between 3.6 cm and 1.3 cm. There is no detection of the source in the CORNISH survey or the catalog of \citet{beck91} which may be due to a combination of sensitivity and spatial filtering by the interferometer. Since the emission measure is less than $2.7 \times 10^8$ pc cm$^{-6}$, this source is classified as a UC \ion{H}{ii} region. The uv-tapered images also led to the identification of a nearby UC \ion{H}{ii} region (J2000 coordinates: 19$^h$ 03$^m$ 46$^s$.0, 5$^\circ$ 40$'$ 42$''$) with 3.6 cm, 1.3 cm and 6.9 mm fluxes of 4.3, 3.3 and 3.4 mJy respectively.

{\it 40.28--0.22} -- The 1.3 cm flux density in this source is $0.4 \pm 0.2$ mJy, while the 6.9 mm flux density is $6.0 \pm 1.2$ mJy, which gives a spectral index of $4.1^{+1.4}_{-0.9}$, which is inconsistent with free-free emission. This suggests that dust emission contributes to the 6.9 mm flux density. One can obtain an estimate of the contribution of dust emission by fitting the 1.2 mm, 870 $\mu$m and 24 $\mu$m flux densities with a grey-body of the form $B_\nu(T_d)~(1-e^{-\tau_\nu})$ where $T_d$ is the dust temperature and $\tau_\nu$ is the optical depth and is given by $\tau_0(\nu/\nu_0)^\beta$, where $\beta$ is the dust emissivity and is assumed to be 2. It should be noted that previous work (e.g. \citealt{mini05}) typically requires more than one temperature component to fit the SED at wavelengths shorter than $\sim 50~\mu$m. However, since we have only three data points, it is not possible to carry out a multi-temperature fit in our case. The consequence of using a single temperature model would be to overestimate the dust temperature and consequently the contribution of dust emission to the 6.9 mm emission. We obtain a dust temperature of 47 K from this fit, and estimate the contribution to the 6.9 mm flux density to be 3.4 mJy. Taking this into account, the free-free spectral index is $2.8 \pm 1.6$, which is marginally consistent with optically thick free-free emission.

\begin{figure}
\centering
\includegraphics[width=0.45\textwidth]{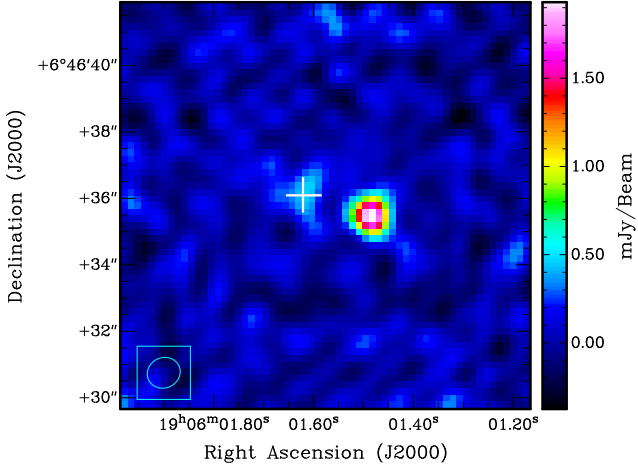}
\caption{The 1.3 cm VLA image of the source 40.62--0.14. The red cross marks the location of the methanol maser.}
\label{g40.62k}
\end{figure}

{\it 40.62--0.14} -- The 1.3 cm map of this source (Fig. \ref{g40.62k}) shows two sources, one of which is coincident with the maser location. The continuum associated with the maser source is not detected at 3.6 cm, and has 1.3 cm and 6.9 mm flux densities of $0.9 \pm 0.3$ mJy and $3.9 \pm 1.7$ mJy respectively. The other stronger source, at J2000 coordinates (19$^h$ 06$^m$ 01$^s$.48, 6$^\circ$ 46$'$ 35$''$.5), has 3.6 cm, 1.3 cm and 6.9 mm flux densities of $3.0 \pm 0.6$ mJy, $2.0 \pm 0.2$ mJy and $3.8 \pm 1.3$ mJy respectively. The two sources are unresolved in the MAMBO and LABOCA maps. A grey-body fit to the 1.2 mm, 870 $\mu$m and 21 $\mu$m data (the latter being from MSX since the MIPS data is completely saturated) gives the contribution of dust to the 6.9 mm emission as 1.7 mJy. Taking into account the error bars, the maser source is consistent with a HC \ion{H}{ii} region, while the nearby source appears to be a UC \ion{H}{ii} region. The latter is not detected in CORNISH (3$\sigma$ limit of 0.9 mJy), which suggests that the emission measure is between $8.5 \times 10^7$ and $2.7 \times 10^8$ pc cm$^{-6}$, similar to that of 38.66+0.08. The GLIMPSE source in Table \ref{table3} is coincident with the maser source.

{\it 41.08--0.13} -- There are two possible MIPSGAL counterparts for this source as indicated in Table \ref{table1}. The infrared flux densities of the two sources are given in Table \ref{table3} by the suffixes `A' and `B' respectively (`A' referring to the position in Table \ref{table1}, and `B' referring to the source in the footnote). Based on the properties of the GLIMPSE source, the source `A' is more likely to be the infrared counterpart for the maser.

{\it 41.12--0.22} -- There are two possible GLIMPSE point sources (labeled `A' and `B' in Table \ref{table3}) that could be mid-infrared counterparts to the maser. The separation of the `A' and `B' sources from the maser are 0.7$''$ and $1.4''$ respectively, while the separation between the two sources and the 24 $\mu$m MIPS source are 2.0$''$ and 2.1$''$ respectively.

{\it 42.43--0.26} -- This is the strongest source at centimeter wavelengths in our sample, with a 3.6 cm flux density of 59.5 mJy. The 6 cm flux density measured by the CORNISH survey is 105.4 mJy, which is much higher than what is expected from extrapolating the 3.6 cm flux density using optically thin free-free emission. However, this is most probably due to the complex and extended structure in the source which can be seen in the 3.6 cm and 1.3 cm maps. Since the emission is optically thin to at least 6 cm, the emission measure is lower than $8.5 \times 10^7$ pc cm$^{-6}$, and is a UC \ion{H}{ii} region. The relatively evolved nature of the source is also evident in this being the only source detected in the 2MASS survey at $J$--band.

{\it 42.70--0.15} -- This source is detected at 1.3 cm at the 8$\sigma$ level though the signal to noise ratio in the integrated flux density is less. The 3.6 cm image shows a 3$\sigma$ source with a peak intensity of 0.4 mJy/beam located 0.6$''$ away from the 1.3 cm source. However, it is not clear whether this source is real, and it is not possible to obtain a good estimate of the integrated intensity of the source. Consequently, this is reported as a non-detection at 3.6 cm in Table \ref{table2}. Assuming that the 3.6 cm source is real and is unresolved, the source would be classified as a hypercompact \ion{H}{ii} region. 

\section{Discussion}
Using eq. (1) for the optical depth of free-free emission under the Altenhoff approximation, an optical depth of unity at 1.3 cm and 6.9 mm correspond to emission measures of $2.1 \times 10^9$ and $8.4 \times 10^9$ pc cm$^{-6}$ respectively. Hence, as mentioned in \S\ref{indi}, the two sources that have continuum emission that is optically thick between 1.3 cm and 6.9 mm (40.28--0.22 and 40.62--0.14) can be classified as hypercompact \ion{H}{ii} regions. Three sources, 38.66+0.08, 39.39--0.14 and 42.43--0.26, are ultracompact \ion{H}{ii} regions, while the source 42.70--0.15 is potentially a hypercompact \ion{H}{ii} region.

The lack of centimeter wave counterparts for the majority of the sample is consistent with the results from previous work. This could arise from two scenarios: the methanol masers could be associated with very early phases of massive star formation that had their \ion{H}{ii} regions quenched by rapid accretion. Alternatively, the masers could arise around intermediate mass stars whose ionizing flux is too weak to result in detectable \ion{H}{ii} regions. 

To try to distinguish between the two scenarios, we adopt two approaches. First, we use the centimeter wave data to obtain upper limits on the sizes of any undetected \ion{H}{ii} regions and the ionizing radiation of the central star. Assuming that any undetected \ion{H}{ii} region is optically thick at frequency, $\nu$, and that the electron temperature in the \ion{H}{ii} region is $T_e = 10^4$ K, the solid angle of the source, $\Omega_s$ is related to the flux density, $S_\nu$ by 
\begin{equation}
S_\nu = B_\nu(T_e) \Omega_s
\end{equation}
where $B_\nu(T)$ is the black body function. Using the $3\sigma$ flux density limits at 1.3 cm ($\nu = 22.46$ GHz), we calculate the limits on the source size, which are tabulated in Table \ref{table4}. Further, we have a constraint on the emission measure, which to first order can be taken to be $n_e^2 l$ where $n_e$ is the electron density and $l$ is the size of the source. We then use ionization equilibrium to calculate limits on the ionizing flux of the central star, the results being tabulated in Table \ref{table4}.

\begin{table}
\caption{Constraints on the sizes of ionized regions and ionizing flux from the central star for sources with no detections at 1.3 cm. The columns show the source name, the limits on the source size, $l$ in milliparsecs and AU, and the limit on the ionizing flux divided by the optical depth of free-free emission at 1.3 cm.}
\label{table4}
\centering
\begin{tabular}{lccc}
\hline \hline
Source & $l$ & $l$ & ${\cal N}_*/\tau_{1.3}$ \\
 & (mpc) & (AU) & $10^{45}$ photons s$^{-1}$ \\
\hline
38.92-0.36 & 0.47 &  98 & 2.5 \\
39.54-0.38 & 0.41 &  84 & 1.9 \\
40.94-0.04 & 0.45 &  93 & 2.3 \\
41.08-0.13 & 0.38 &  78 & 1.6 \\
41.12-0.11 & 0.45 &  93 & 2.3 \\
41.12-0.22 & 0.39 &  81 & 1.7 \\
41.16-0.20 & 0.39 &  81 & 1.7 \\
41.23-0.20 & 0.39 &  81 & 1.7 \\
41.27+0.37 & 0.52 & 107 & 3.0 \\
41.34-0.14 & 0.52 & 108 & 3.1 \\
41.58+0.04 & 0.52 & 107 & 3.0 \\
42.03+0.19 & 0.50 & 103 & 2.8 \\
42.30-0.30 & 0.47 &  98 & 2.5 \\
43.08-0.08 & 0.51 & 104 & 2.9 \\
\hline
\end{tabular}
\end{table}

The limits on the ionizing flux from the central object are not inconsistent with the presence of massive stars since the optical depth of free-free emission at 1.3 cm ($\tau_{1.3}$ in Table \ref{table4}) is unknown. However, the limits on the source size show that any ionized region must be confined to a volume that is comparable to that of the solar system (up to the Kuiper belt). As mentioned previously, these results could also be explained by the central objects being intermediate mass stars.

It is worth pointing out that the HC \ion{H}{ii} regions associated with CRL 2136, W33 A, NGC 2591, NGC 7538 IRS9 \citep{ment04, van05}, which all have bona fide associated class II CH$_3$OH masers, would not or only marginally have been detected at the flux density limits of our VLA observations.  Consequently, the sizes that these authors determine for the above sources (80, 150, 20, and 50 AU, respectively) are all at or lower than the upper limits on sizes we present in Table \ref{table4}.

An alternate approach to the problem is to fit the infrared flux densities using models of young stellar objects which provides constraints on the physical parameters of the sources. We used the SED fitter of \citet{robi07}, which uses a grid of 200,000 precomputed model SEDs spanning a wide range of evolutionary stages for different stellar masses (assuming that stars form by accretion through a disk and envelope). The fitting procedure involves interpolation of the model fluxes to the apertures used to perform the photometry, scaling them to a number of distances between $d_{\mathrm{min}}$ and $d_{\mathrm{max}}$ (calculated from the uncertainties in the distance determination), followed by fitting to the data with the visual extinction, $A_V$, being a free parameter (allowed to vary between 0 and 100 magnitudes in our fitting). The extinction model used by \citet{robi07} is the method of \citet{kim94} modified for the mid-infrared extinction properties derived by \citet{inde05}.

To account for possible calibration errors between different data sets, we set the minimum uncertainty in the flux densities to be 10\%. For the millimeter and submillimeter data, the aperture sizes were set to the source sizes measured by ``JMFIT''. The measured source sizes for the 1.2 mm data typically ranged from 10 to 40$''$, while for the 870 $\mu$m data, the sizes were 20 to 40$''$. The visual extinction was set to vary between 0 and 100 magnitudes. The SED fitter provides the parameters of the models that fit the data in order of increasing $\chi^2$ goodness of fit. We calculate the mean and standard deviation of the stellar mass ($M_*$), temperature ($T_*$), radius ($R_*$), envelope accretion rate ($\dot{M}$), age, total luminosity ($L_{tot}$), visual extinction ($A_V$), and inclination to the line of sight ($i$) from the models using weighting in accordance with the $\chi^2$ probability distribution. The standard deviation is typically asymmetric about the mean, and in some cases unrealistically small when few models fit the data and the difference in $\chi^2$ between successive models is large. The results of this study are summarized in Table \ref{table5}, and the model fits are shown in Figure \ref{sedfits}. We do not show any disk parameters since the models are dominated by the envelope and the disk mass is small compared to the mass of the central star.

\begin{table*}
\caption{Physical parameters of the young stellar objects associated with 6.7 GHz methanol masers from modeling their SEDs. The columns show the source name, age, mass, temperature and radius of the central object, envelope accretion rate, total luminosity, visual extinction to the source and model inclination to the line of sight. Note that the uncertainties are calculated solely from the $\chi^2$ statistics of different models that fit the data and do not take into account uncertainties from source multiplicity or other systematic uncertainties in the underlying model assumptions.}
\label{table5}
\centering
\begin{tabular}{lcccccccc}
\hline \hline
Source & Age & $M_*$ & $T_*$ & $R_*$ & $\dot{M}$ & $L_{tot}$ & $A_V$ & $i$ \\
 & ($10^3$ yr) & ($M_\odot$) & (K) & ($R_\odot$) & ($10^{-3}~M_\odot$/yr) & ($10^3~L_\odot$) & (mag) & ($^\circ$) \\
\hline
38.66+0.08 & $9.5^{+23.3}_{-0.3}$ & $16.8^{+0}_{-1.3}$ & $8600^{+16\,100}_{-200}$ & $62^{+1}_{-52}$ & $6.0^{+0}_{-0.2}$ & $18^{+16}_{-0}$  & $55^{+17}_{-0}$ & $18$ \\
38.92-0.36 & $4.9^{+0}_{-3.3}$ & $22.7^{+0.2}_{-9.0}$ & $8900^{+100}_{-4700}$ & $87^{+53}_{-1}$ & $7.7^{+0.1}_{-5.4}$ & $42^{+10}_{-37}$  & $26^{+48}_{-1}$  & $41^{+1}_{-23}$ \\
39.39-0.14 & $50^{+12}_{-11}$ & $10.9^{+0.6}_{-0.4}$ & $13\,300^{+1200}_{-5300}$ & $18^{+20}_{-4}$ & $2.6^{+0.5}_{-0.3}$ & $8.1^{+0.8}_{-0.4}$  & $21 \pm 4$  & $51^{+6}_{-8}$ \\
39.54-0.38 & $27^{+2}_{-9}$ & $7.8^{+1.2}_{-0.2}$ & $4600^{+700}_{-0}$ & $38^{+15}_{-0}$ & $1.7^{+0.8}_{-0.07}$ & $0.61^{+1.72}_{-0.04}$  & $26^{+5}_{-24}$ & $31^{+1}_{-13}$ \\
40.28-0.22 & $4.1^{+0}_{-0.3}$ & $38.4^{+0}_{-1.2}$ & $15\,400^{+0}_{-1400}$ & $65^{+25}_{-0}$ & $9.8^{+0}_{-2.3}$ & $210^{+0}_{-30}$  & $5$  & $32^{+0}_{-14}$ \\
40.62-0.14 & $30^{+50}_{-0}$ & $19.4^{+4.2}_{-0}$ & $32\,300^{+5300}_{-0}$ & $7^{+0}_{-1}$ & $2.3^{+0}_{-0.3}$ & $62^{+13}_{-0}$  & $31^{+0}_{-7}$  & $32^{+10}_{-0}$ \\
41.08-0.13A & $5.9^{+0}_{-2.3}$ & $16.4^{+3.8}_{-0}$ & $5300^{+0}_{-600}$ & $150^{+90}_{-0}$ & $7.3^{+0.7}_{-0}$ & $17^{+10}_{-0}$  & \tablefootmark{a}  & $32^{+24}_{-0}$ \\
41.08-0.13B & $1.2^{+1.5}_{-0}$ & $27.1^{+0.2}_{-4.9}$ & $4200^{+300}_{-0}$ & $290^{+0}_{-20}$ & $7.9^{+1.8}_{-0.1}$ & $23^{+3}_{-0.1}$  & \tablefootmark{a}  & 32 \\
41.12-0.11 & $6.5^{+1.6}_{-3.7}$ & $27.9^{+11.1}_{-1.6}$ & $14\,000^{+2800}_{-6500}$ & $100^{+370}_{-57}$ & $6.2^{+0.5}_{-1.1}$ & $95^{+5}_{-19}$  & $37^{+2}_{-1}$  & 18 \\
41.12-0.22A & $7.4^{+23.4}_{-2.8}$ & $21.6^{+1.4}_{-3.5}$ & $9730^{+13\,100}_{-2400}$ & $100^{+120}_{-37}$ & $7.0^{+0.9}_{-2.2}$ & $38^{+6}_{-13}$  & $17^{+6}_{-12}$  & $34^{+8}_{-2}$ \\
41.12-0.22B & $18^{+15}_{-14}$ & $18.8 \pm 3.3$ & $16\,000^{+8700}_{-8300}$ & $68^{+75}_{-56}$ & $6.6^{+1.0}_{-0.7}$ & $34^{+6}_{-4}$  & $15^{+9}_{-8}$  & $37 \pm 5$ \\
41.23-0.20 & $2.8^{+1.0}_{-1.1}$ & $12.2^{+6.7}_{-0.5}$ & $4400^{+300}_{-100}$ & $110^{+31}_{-18}$ & $4.8^{+0.6}_{-0.4}$ & $4.7^{+22.3}_{-1.4}$  & $50^{+10}_{-15}$  & $19^{+16}_{-1}$ \\
41.27+0.37 & $2.7^{+2.0}_{-0.1}$ & $16.2^{+0.3}_{-0.2}$ & $4300^{+400}_{-0}$ & $140^{+6}_{-4}$ & $3.9^{+0.9}_{-1.5}$ & $6.4^{+0.6}_{-0.2}$  & $7 \pm 1$  & 18 \\
41.34-0.14 & $6.4^{+26.6}_{-2.0}$ & $11.5^{+4.7}_{-0.3}$ & $4900^{+8400}_{-500}$ & $84^{+15}_{-44}$ & $5.1^{+0.3}_{-2.7}$ & $4.0^{+13.4}_{-0.7}$  & $34^{+9}_{-31}$  & $20^{+15}_{-2}$ \\
42.03+0.19 & $57^{+0}_{-7}$ & $21.3^{+3.0}_{-0}$ & $36\,300^{+1800}_{-100}$ & $6^{+8}_{-0}$ & $5.2^{+0}_{-2.8}$ & $55^{+22}_{-0}$  & $24^{+1}_{-0}$  & $31^{+1}_{-13}$ \\
42.30-0.30 & $2.2^{+1.5}_{-0.1}$ & $13.2^{+3.2}_{-0.3}$ & $4300^{+8800}_{-0}$ & $120^{+84}_{-0}$ & $5.2^{+2.5}_{-0.4}$ & $4.6^{+3.5}_{-0.1}$  & $26^{+2}_{-10}$  & $18^{+14}_{-0}$ \\
42.43-0.26 & $47^{+5}_{-0}$ & $16.4^{+1.2}_{-0}$ & $30\,900^{+2100}_{-0}$ & $5.8^{+0}_{-0.2}$ & $3.7^{+1.6}_{-0}$ & $29^{+6}_{-0}$  & $10^{+0}_{-4}$  & $18^{+14}_{-0}$ \\
42.70-0.15 & $16^{+6}_{-4}$ & $15.7^{+2.0}_{-0.7}$ & $10\,900^{+4600}_{-2200}$ & $41^{+31}_{-9}$ & $4.7^{+1.0}_{-1.8}$ & $17^{+23}_{-2}$  & $66^{+15}_{-6}$  & $19^{+15}_{--}$ \\
43.08-0.08 & $4.3^{+6.9}_{-0.7}$ & $24.5^{+2.9}_{-0.3}$ & $8600^{+11\,200}_{-1300}$ & $140^{+140}_{-100}$ & $7.3^{+0.9}_{-0.1}$ & $61^{+38}_{-5}$  & $27^{+8}_{-1}$  & $33^{+12}_{-2}$ \\
\hline
\tablefoottext{a}{$A_V$ not well constrained.}
\end{tabular}
\end{table*}

\begin{figure*}
\centering
\includegraphics[height=0.945\textheight]{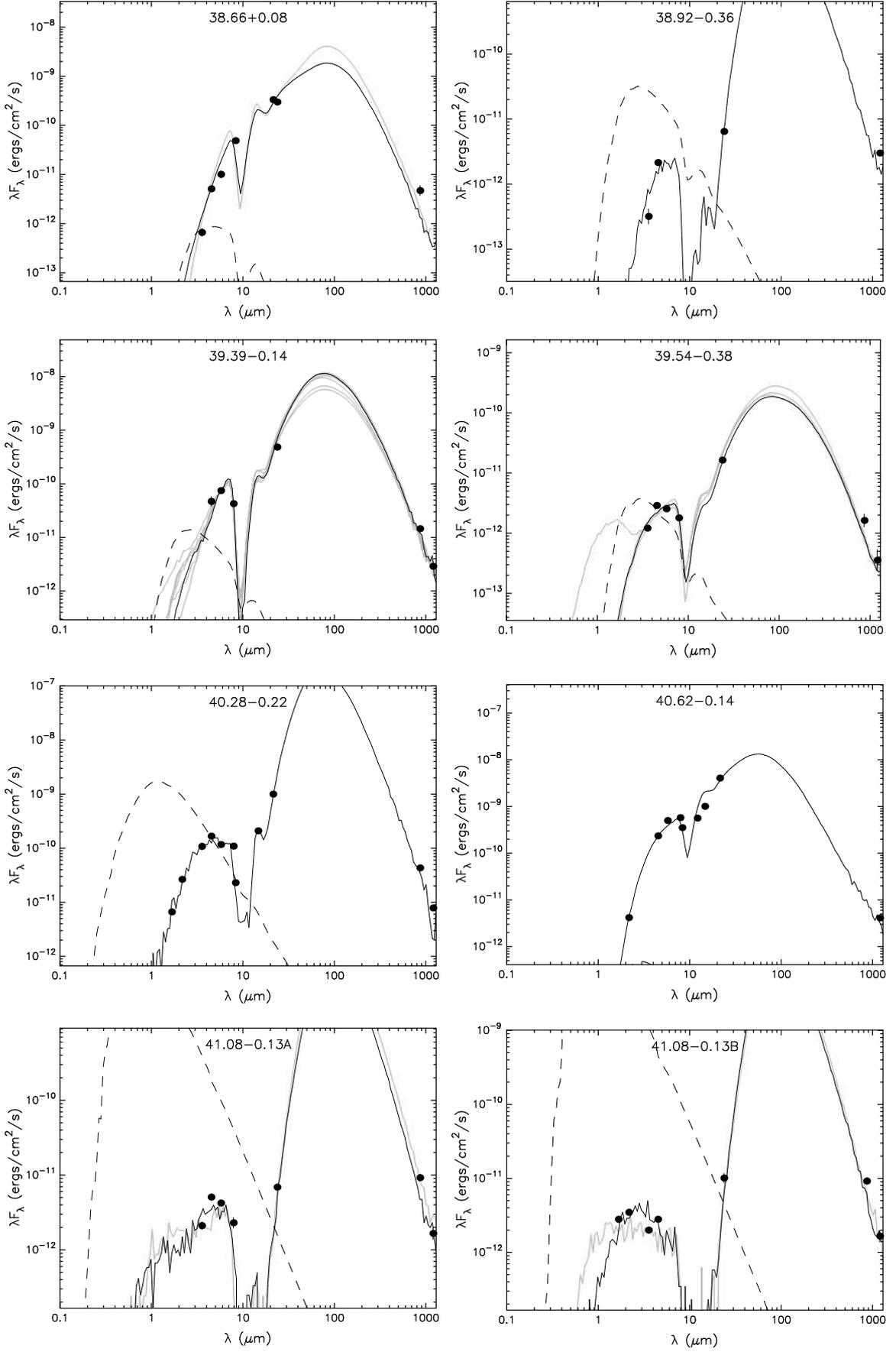}
\caption{Fits to the SEDs of 6.7 GHz methanol masers. The solid line shows the best fitting model, while the grey lines show other models whose $(\chi^2 - \chi^2_{best})/\mathrm{datapoint} < 3$. The dashed line shows the stellar photosphere corresponding to the central star in the best fitting model in the absence of circumstellar dust, but including interstellar extinction.}
\label{sedfits}
\end{figure*}

\setcounter{figure}{2}
\begin{figure*}
\centering
\includegraphics[height=0.97\textheight]{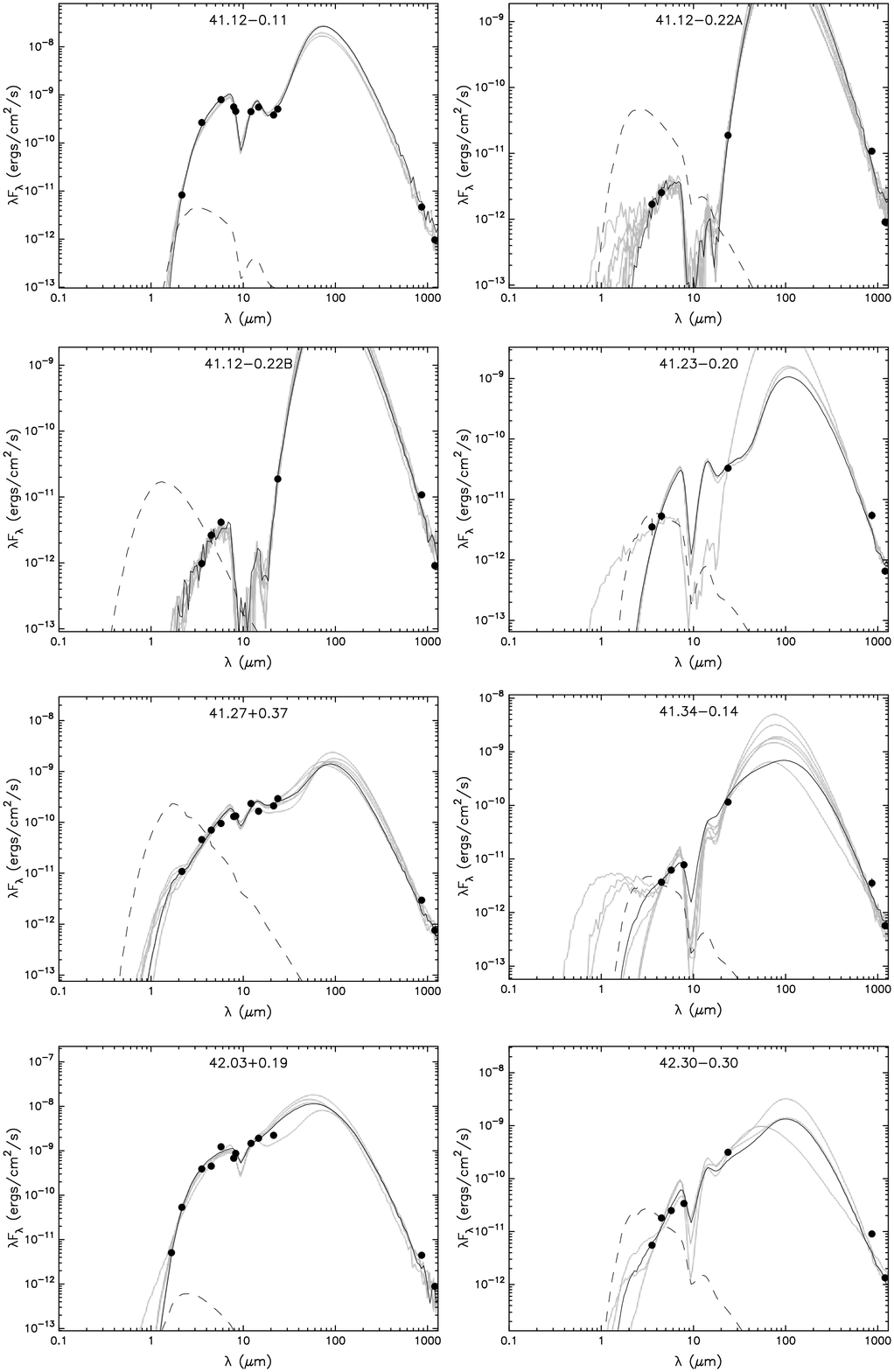}
\caption{Continued.}
\end{figure*}

\setcounter{figure}{2}
\begin{figure*}
\centering
\includegraphics[width=0.85\textwidth]{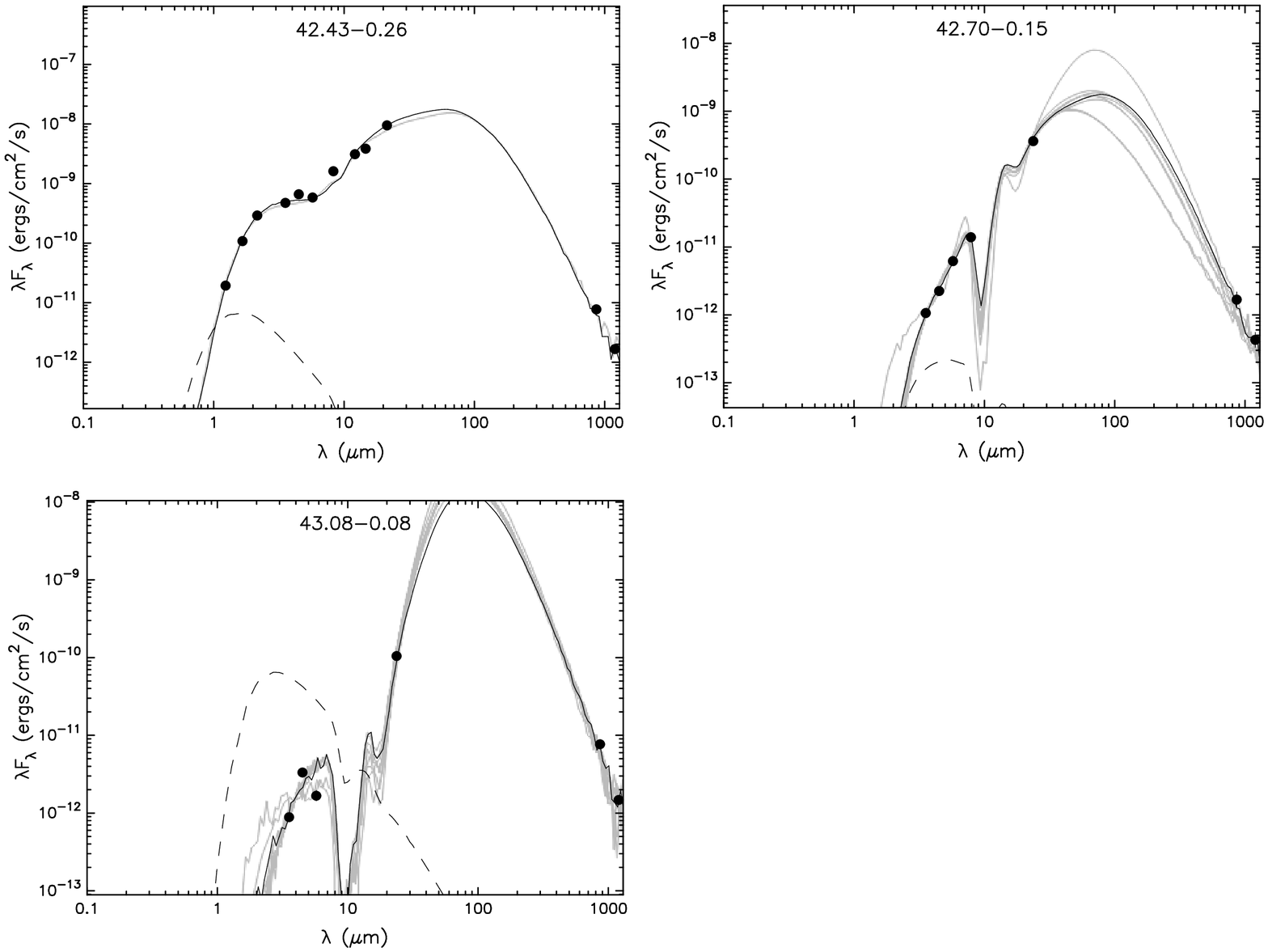}
\caption{Continued.}
\end{figure*}

Three sources are not included in Table \ref{table5}. The sources 41.16--0.20 and 41.58+0.04 do not have 1.2 mm or 870 $\mu$m flux densities, and in the absence of at least one of these two flux densities, SED fits are poorly constrained. In addition, a large number of models fit the data for source 40.94--0.04, and so meaningful parameters cannot be deduced by this method. As noted in \S\ref{indi} and Table \ref{table3}, there are two possible infrared counterparts for 41.08--0.13 and 41.12--0.22. Hence, both possible SEDs have been modeled and the results indicated by suffixes `A' and `B' in Table \ref{table5} and Figure \ref{sedfits}.

It is to be noted that uncertainties shown in Table \ref{table5} only take into account the $\chi^2$ statistics of multiple models fitting the data, and do not show the systematic uncertainties from the underlying model assumptions. For example, model parameters such as stellar mass and age are derived from the luminosity and temperature of the star using evolutionary tracks \citep{robi08}. Consequently, uncertainties in the evolutionary tracks translate to uncertainties in the masses and ages. These uncertainties have not been quantified and are not shown in Table \ref{table5}. Further, the envelope accretion rate shown in Table \ref{table5} is derived from the density profile measured from the SED through the assumption of a free-fall rotational collapse model \citep{robi08}. Since the millimeter and submillimeter data (which primarily trace the envelope) have relatively poor angular resolution, source multiplicity potentially leads to over-estimation of the envelope accretion rates. 

Keeping the caveats above in mind, Table \ref{table5} shows that the YSO models that fit the infrared and submillimeter data points are rapidly accreting massive stars. The central stars have masses larger than $\sim 8~M_\odot$, and considering the rapid accretion, the final mass of the central star is likely to be much higher. In the case of the source 41.58+0.04, the upper limit at 1.2 mm corresponds to an isothermal mass of between 38 and 87 $M_\odot$ depending on the prescription of dust opacity \citep[e.g.][]{osse94,hild83} assuming the dust temperature to be 30 K, which is adequate to host a massive star (although the constraint is much tighter if one uses the $1\sigma$ rather than $3\sigma$ limit).

The accretion rates and stellar temperatures in Table \ref{table5} give further confirmation of the results at centimeter wavelengths. The stellar temperatures are mostly too cool to form \ion{H}{ii} regions (which is consistent with their young age and rapid accretion; e.g. \citealt{hoso09}). Moreover, as pointed out by \citet{walm95}, the ``critical accretion rate'' required to confine the \ion{H}{ii} region to a volume close to the stellar surface is $10^{-4}$ to $4 \times 10^{-6}~M_\odot$ yr$^{-1}$ for stars of spectral type O5 ($\sim 60~M_\odot$) to B0 ($\sim 17~M_\odot)$ respectively. The accretion rates deduced from the SED fitting are well above the critical accretion rate. This result is unlikely to change even accounting for source multiplicity, which is common in the context of massive star forming regions. However, this calculation assumes that the accretion is spherical, which is not true in practice. This might explain why some of the sources are seen as ultracompact and hypercompact \ion{H}{ii} regions in spite of having accretion rates that would prohibit such regions from forming.

We cannot deduce a minimum mass for the central objects associated with 6.7 GHz methanol masers since our sample size is relatively small. Our work does not rule out the possibility of a small number of masers being associated with stars with spectral type later than B3. However, accepting the assumption that all the methanol masers in our sample are massive YSOs (the mystery of 41.58+0.04 not withstanding), it is interesting to see that only six sources out of twenty are associated with HC \ion{H}{ii} and UC \ion{H}{ii} regions (although the absence of HC \ion{H}{ii} regions could be due to sensitivity limitations as discussed earlier). \citet{wals98} suggested that methanol masers are destroyed as a UC \ion{H}{ii} region develops. However, there are sources such as W3(OH) \citep{ment92} and NGC 7538 IRS1 \citep{mini00,mosc09} where methanol masers are associated with strong and well developed UC \ion{H}{ii} regions (sizes $> 1000$~AU). Furthermore, \citet{hart95} argue that OH and CH$_3$OH can coexist in the envelope of a UC \ion{H}{ii} region. 

One might argue that the OH maser switches on at a later evolutionary stage and lasts longer. This appears to observationally confirmed: many methanol masers do not have associated OH masers. Although this may, again, be a selection effect caused by the 6.7 GHz methanol maser's intensity to be on average $\approx 7$ times higher than the strongest transition of the OH maser \citep{casw95}, it might indeed be the case. Both the high OH and CH$_3$OH abundances are the result of frozen out OH and H$_2$O being evaporated off dust grain mantles by heating from the igniting protostar. However, the production of OH by photodissociation of H$_2$O requires ultraviolet radiation, which necessitates the presence of an UC \ion{H}{ii} region. 

In addition, the estimated lifetime of the methanol maser emission is only a few $\times 10^4$~yr \citep{vanw05}. These would argue against the suggestion that the development of the UC \ion{H}{ii} region destroys the methanol maser emission. However, our data does show that the likelihood of the presence of maser emission decreases as the massive star evolves into an \ion{H}{ii} region. Thus, our work would be consistent with a statistical statement that 6.7 GHz methanol masers are most likely to be associated with massive stars prior to the formation of a HC \ion{H}{ii} region, and disappear during the UC \ion{H}{ii} phase.

As mentioned previously, one of the big caveats in this study is the sources may resolve into multiple components at high angular resolution. This is especially so for the millimeter and submillimeter data which have much poorer resolution compared to other data. While the fitting procedure takes into account the aperture used for calculating the flux density of the source, the assumption is that the dusty envelope seen in the millimeter and submillimeter data harbors a single massive YSO. Data at high angular resolution is required to determine the severity of the effect of multiplicity in these results. Observations with existing and upcoming submillimeter interferometers will be able to address this question. A more complete description of the limitations of the SED modeling employed here can be found in \citet{robi08}.

\section{Conclusions}

We have obtained SEDs of a complete sample of 20 6.7 GHz methanol masers from centimeter to near-infrared wavelengths by combining new observations with existing data. Centimeter wave continuum is not seen in 14 sources, while three sources each are associated with HC \ion{H}{ii} and UC \ion{H}{ii} regions respectively. The upper limits on the sizes of any ionized region around sources with no radio continuum are consistent with such regions being confined close to the stellar surface. Modeling of the SED from millimeter to near-infrared wavelength strongly suggests that the young stellar objects associated with the masers are rapidly accreting young massive stars. Although the relatively small size of our sample does not allow us to put any constraints on the minimum mass of the central star of the YSOs associated with the masers, this work shows that most 6.7 GHz methanol masers are associated with massive stars prior to the formation of a hypercompact \ion{H}{ii} region.

\begin{acknowledgements}
We thank R. Zylka for help in reducing the MAMBO data, and F. Schuller and A. Belloche for help in reducing the LABOCA data. We also thank S. Kurtz and J. Williams for insightful discussions. This publication makes use of data products from the Two Micron All Sky Survey, which is a joint project of the University of Massachusetts and the Infrared Processing and Analysis Center/California Institute of Technology, funded by the National Aeronautics and Space Administration and the National Science Foundation. YX was supported by the Chinese NSF through grants NSF 10673024, NSF 10733030, NSF 10703010 and NSF 10621303, and NBRPC (973 Program) under grant 2007CB815403. This work was supported in part by the Jet Propulsion Laboratory, California Institute of Technology. This research has made use of NASA's Astrophysics Data System.
\end{acknowledgements}

\bibliographystyle{aa}
\bibliography{14937refs}

\Online

\begin{figure*}
\centering
\includegraphics[width=\textwidth]{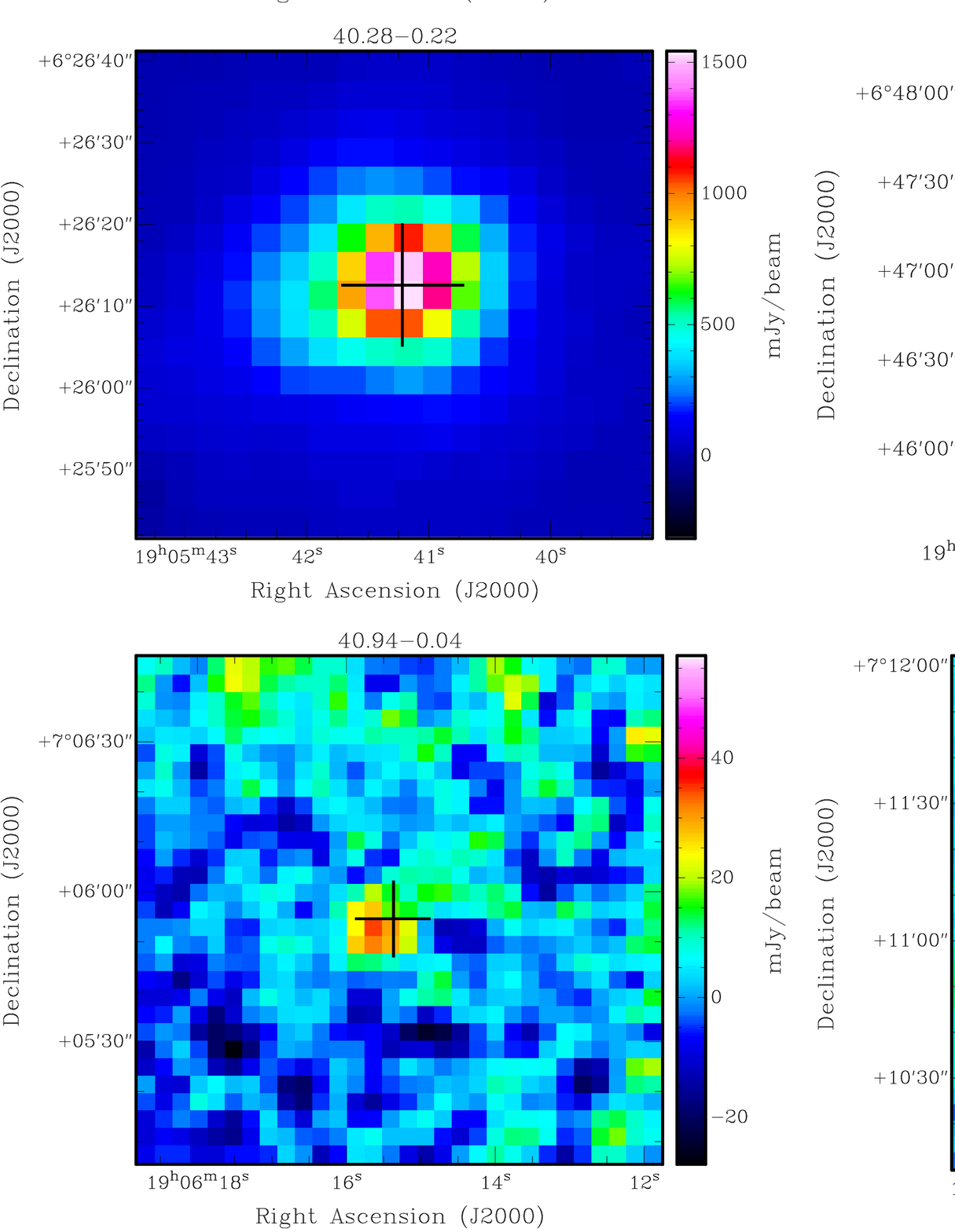}
\caption{The 1.2 mm MAMBO images towards the 6.7 GHz methanol masers in our sample. The black cross shows the location of the methanol maser. The source 38.66+0.08 is not shown for reasons indicated in \S\ref{indi}, while 38.92--0.36 is shown in Fig. \ref{g38.92l}.}
\label{mambomaps}
\end{figure*}

\setcounter{figure}{3}
\begin{figure*}
\centering
\includegraphics[width=\textwidth]{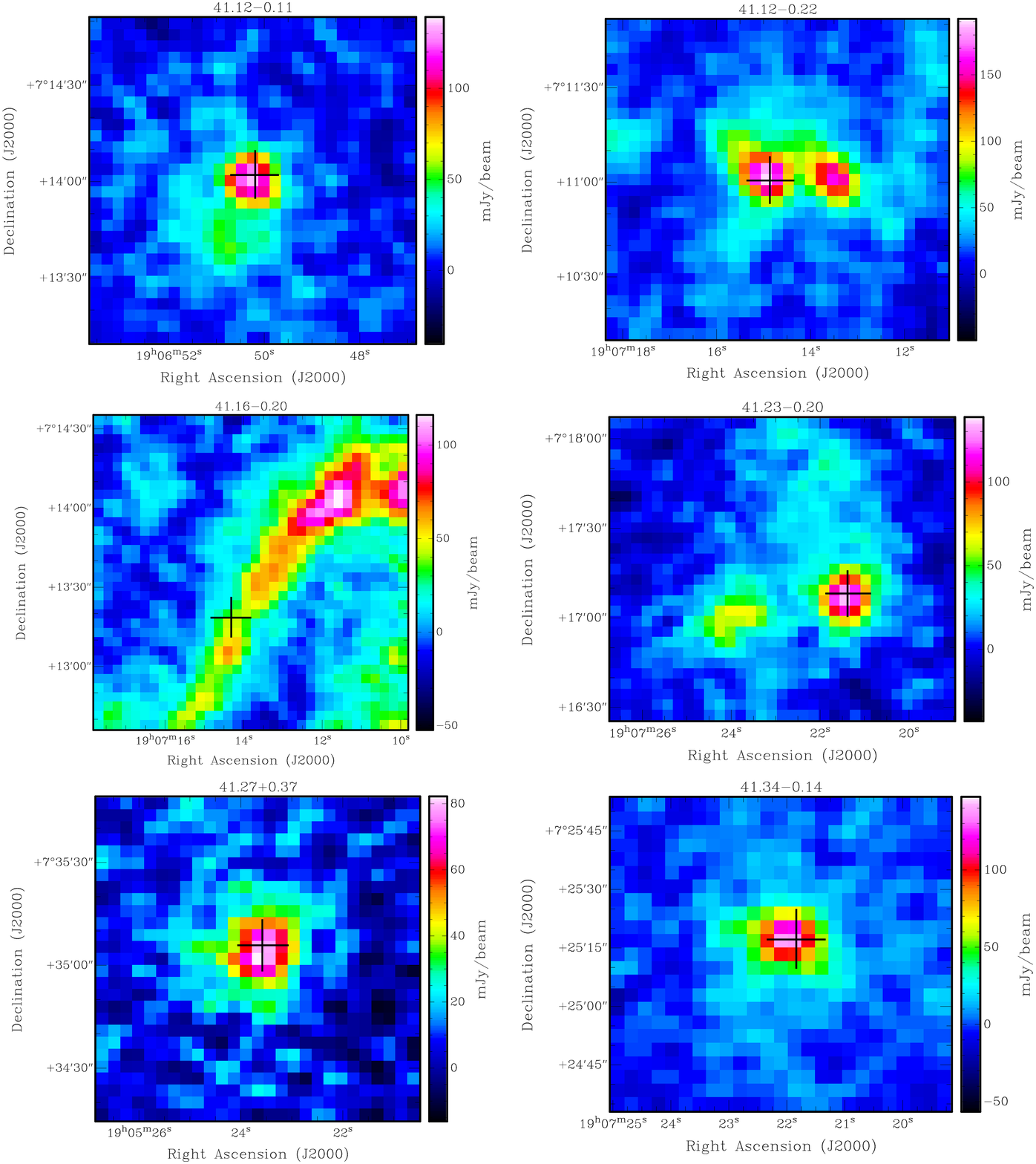}
\caption{Continued.}
\end{figure*}

\setcounter{figure}{3}
\begin{figure*}
\centering
\includegraphics[width=\textwidth]{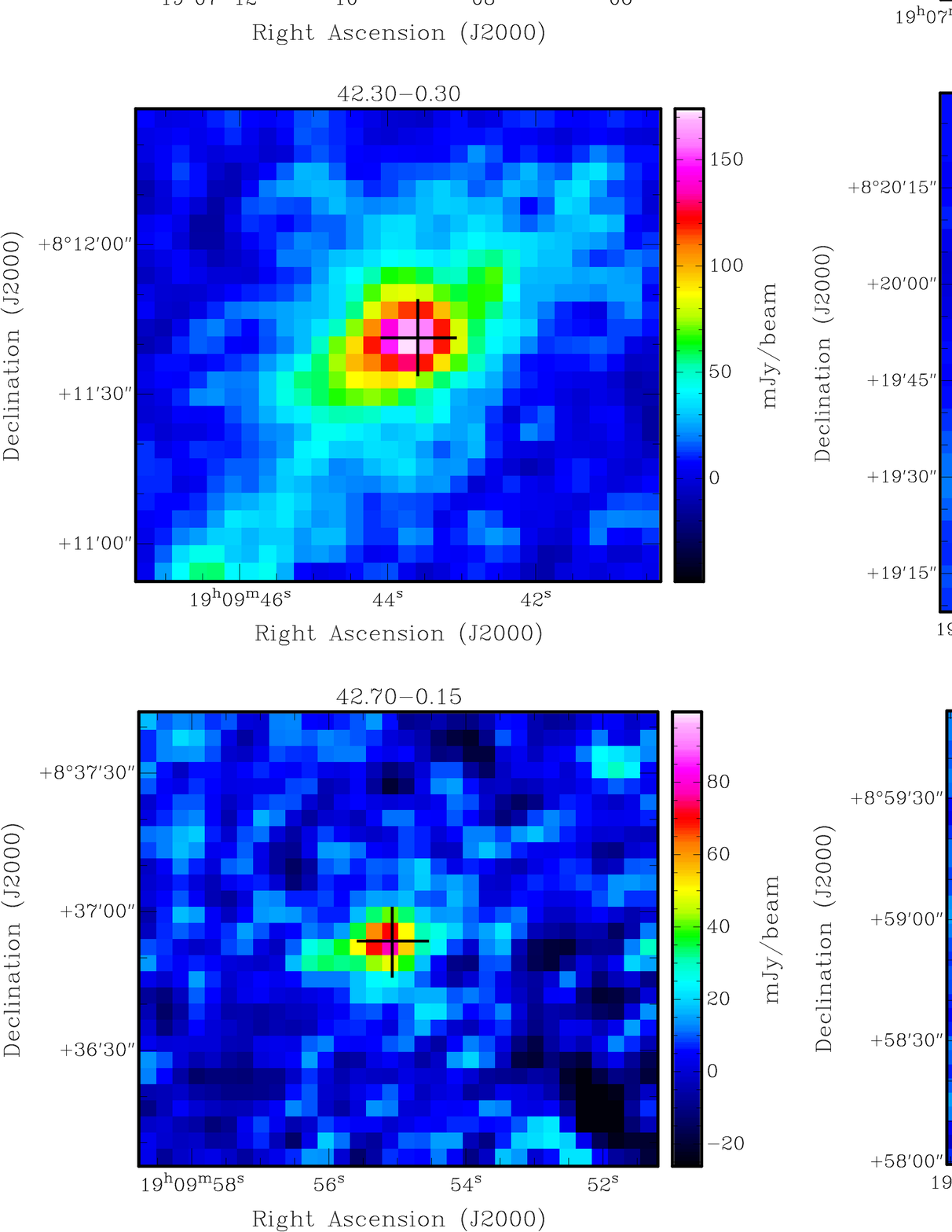}
\caption{Continued.}
\end{figure*}

\end{document}